# RECONSTRUCTING THE INFLATON POTENTIAL

Edward W. Kolb,[1,2] Mark Abney,[2]
Edmund J. Copeland,[3] Andrew R. Liddle,[3] and James E. Lidsey[1]

[1]NASA/Fermilab Astrophysics Center
 Fermi National Accelerator Laboratory, Batavia, IL 60510 USA

[2]Department of Astronomy and Astrophysics, Enrico Fermi Institute
 The University of Chicago, Chicago, IL 60637 USA

[3]School of Mathematical and Physical Sciences
 University of Sussex
 Brighton BN1 9QH, United Kingdom

## INTRODUCTION

Inflation involves a period of rapid growth of the Universe. This is most easily illustrated by considering a homogeneous, isotropic Universe with a flat Friedmann–Robertson–Walker (FRW) metric described by a scale factor $a(t)$. Here, "rapid growth" means a positive value of $\ddot{a}/a = -(4\pi G_N/3)(\rho+3p)$ where $\rho$ is the energy density and $p$ the pressure. It is useful to identify the energy density driving inflation with some sort of scalar "potential" energy density $V > 0$ that is positive, and results in an effective equation of state $\rho \simeq -p \simeq V$, which satisfies $\ddot{a} > 0$. If one identifies the potential energy as arising from the potential of some scalar field $\phi$, then $\phi$ is known as the *inflaton* field.

The prime observational consequences of inflation derive from the stochastic spectra of density (scalar) perturbations and gravitational wave (tensor) modes generated during inflation. Each stretches from scales of order centimeters to scales well in excess of the size of the presently observable Universe. Once within the Hubble radius, gravitational waves redshift away and so their main influence is on the large-scale microwave background anisotropies, such as those probed by COBE [1]. Advanced gravitational wave detectors such as the proposed beam-in-space experiments may be able to detect the gravitational waves on a much shorter (about $10^{14}$cm) wavelength range. The density perturbations are thought to lead to structure formation in the Universe. They produce microwave background anisotropies across a much wider range of angular scales than do the tensor modes, and constraints on the scalar spectrum are also available from the clustering of galaxies and galaxy clusters, peculiar velocity flows and a host of other measurable quantities [2].

Broadly speaking, inflation predicts a very nearly Gaussian spectrum of density perturbations that is *scale dependent*, i.e., the amplitude of the perturbation depends upon the length scale. Such a dependence typically arises because the Hubble expansion rate during the inflationary epoch changes, albeit slowly, as the field driving the expansion rolls towards the minimum of the scalar potential. This implies that the amplitude of the fluctuations as they cross the Hubble radius will be weakly time-dependent.

Within the context of slow-roll inflation, any scale dependence for density perturbations is possible if one considers an arbitrary functional form for the inflaton potential, $V(\phi)$. In this sense, inflation makes no unique prediction concerning the form of the density perturbation spectrum and one is left with two options. Either one can aim to find a deeper physical principle that uniquely determines the potential, or observations that depend on $V(\phi)$ can be employed to limit the number of possibilities. One such observation is the amplitude of the tensor perturbations produced by inflation.

Recently, we provided a formalism which allows one to reconstruct the inflaton potential $V(\phi)$ directly from a knowledge of these spectra [3]. This developed an original but incomplete analysis by Hodges and Blumenthal [4]. An important result that follows from our formalism is that knowledge of the scalar spectrum alone is insufficient for a unique reconstruction. Reconstruction from only the scalar spectrum leaves an arbitrary integration constant, and since the reconstruction is nonlinear, different choices of this constant lead to different functional forms for the potential. A minimal knowledge of the tensor spectrum, say its amplitude at a single wavelength, is sufficient to lift this degeneracy.

The most ambitious aim of reconstruction is to employ observational data to deduce the inflaton potential over the range corresponding to microwave fluctuations and large-scale structure, although at present the observational situation is some way from providing the quality of data that this would require [3].

In this talk I will discuss the promise of potential reconstruction assuming one knows 1) the amplitude of the tensor spectrum at one point from microwave background fluctuations, presumably on quadrupole scales corresponding to $3000h^{-1}$Mpc, and 2) the scalar spectrum from microwave background fluctuations and the large-scale structure investigations from quadrupole scales down to scales of several Mpc.

## PERTURBATIONS FROM SLOW-ROLL INFLATION

We are interested in the perturbations resulting from inflation. The "density" perturbations are usually described in terms of fluctuations in the local value of the mass density. In a Universe with density field $\rho(\mathbf{x})$ and mean mass density $\rho_0$, the density contrast is defined as

$$\delta(\mathbf{x}) = \frac{\delta\rho(\mathbf{x})}{\rho_0} = \frac{\rho(\mathbf{x}) - \rho_0}{\rho_0}. \tag{1}$$

It is convenient to express this contrast in terms of a Fourier expansion:

$$\delta(\mathbf{x}) = A \int \delta_{\mathbf{k}} \exp(-i\mathbf{k}\cdot\mathbf{x}) d^3k, \tag{2}$$

where $A$ is an overall normalization constant, interesting only for those who enjoy keeping track of factors of $2\pi$. What is usually meant by the density perturbation on a scale $\lambda$, $(\delta\rho/\rho)_\lambda$, is related to the square of the Fourier coefficients $\delta_{\mathbf{k}}$:

$$\left(\frac{\delta\rho}{\rho}\right)^2_\lambda \equiv A' \left.\frac{k^3|\delta_k|^2}{2\pi^2}\right|_{\lambda=k^{-1}}, \tag{3}$$

where again we have included an overall normalization constant $A'$. The perturbations are normally taken to be (statistically) isotropic, in the sense that the expectation of $|\delta_{\mathbf{k}}|^2$ averaged over a large number of independent regions can depend only on $k = |\mathbf{k}|$. The dependence of $\delta\rho/\rho$ as a function of $\lambda$ is the spectrum of the density perturbations.

In a spatially flat isotropic Universe the Hubble expansion rate is $H(t) = \dot{a}/a$, and its inverse $H^{-1}(t)$ (the Hubble radius) is the scale beyond which causal processes no longer operate. Of crucial importance is the relative size of a scale $\lambda$ to the Hubble radius. The *physical* length between two points of coordinate separation $d$ is $\lambda(t) = a(t)d$, so that a length scale comoving with the expansion will grow in proportion to $a(t)$. The condition for inflation to occur is precisely the condition for physical scales to grow more rapidly than the Hubble length; that is, for the comoving Hubble radius $H^{-1}/a$ to shrink. Thus, a given scale can start sub-Hubble radius, $\lambda < \lambda_H$, pass outside the Hubble radius during inflation, and finally re-enter the Hubble radius long after inflation. Thus, perturbations can be imparted on a given length scale in the inflationary era as that scale leaves the Hubble radius, and will be present as that scale re-enters the Hubble radius after inflation in the radiation-dominated or matter-dominated era.

Microphysics cannot affect the perturbation while it is outside the Hubble radius, and the evolution of its amplitude is *kinematical*, unaffected by dissipation, the equation of state, instabilities, and the like. However, for super-Hubble-radius sized perturbations one must take into account the freedom in the choice of the background reference space-time, i.e., the gauge ambiguities. As usual when confronted with such a problem, it is convenient to calculate a *gauge-invariant* quantity. For inflation it is convenient to study the Bardeen potential $\zeta$ [5]. In the uniform Hubble constant gauge $\zeta$ is particularly simple. It is related to the background energy density and pressure, $\rho_0$ and $p_0$, and the perturbed energy density $\rho_1$ by $\zeta \equiv \delta\rho/(\rho_0 + p_0)$, where $\delta\rho = \rho_1 - \rho_0$ is the density perturbation.

In the standard matter-dominated (MD) or radiation-dominated (RD) phase, $\zeta$ at Hubble radius crossing is equal (up to a factor of order unity) to $\delta\rho/\rho$. Thus, the amplitude of a density perturbation when it crosses back inside the Hubble radius after inflation, $(\delta\rho/\rho)_{\text{HOR}}$,[1] is given by $\zeta$ at the time the fluctuation crossed outside the Hubble radius during inflation.

As inferred from the adoption of $\zeta$, the convenient specification of the amplitude of density perturbations on a particular scale is when that particular scale just enters the Hubble radius, denoted as $(\delta\rho/\rho)_{\text{HOR}}$. Specifying the amplitude of the perturbation at Hubble radius crossing evades the subtleties associated with the gauge freedom, and has the simple Newtonian interpretation as the amplitude of the perturbation in the gravitational potential. Of course, when one specifies the fluctuation spectrum at Hubble radius crossing, the amplitudes for different lengths are specified at *different* times.

Now let us turn to the scalar field dynamics during inflation. Consider a minimally coupled, spatially homogeneous scalar field $\phi$, with Lagrangian density

$$\mathcal{L} = \partial^\mu \phi \partial_\mu \phi/2 - V(\phi) = \dot{\phi}^2/2 - V(\phi). \tag{4}$$

With the assumption that $\phi$ is spatially homogeneous, the stress-energy tensor takes the form of a perfect fluid, with energy density and pressure given by $\rho_\phi = \dot{\phi}^2/2 + V(\phi)$ and $p_\phi = \dot{\phi}^2/2 - V(\phi)$. The classical equation of motion for $\phi$ is

$$\ddot{\phi} + 3H\dot{\phi} + V'(\phi) = 0, \tag{5}$$

---

[1] The notation "HOR" follows because often in the literature the Hubble radius is referred to (incorrectly) as the horizon.

and the expansion rate in a flat FRW spacetime is given by ($\kappa^2 = 8\pi G_N$)

$$H^2 = \frac{\kappa^2}{3}\left(\frac{1}{2}\dot{\phi}^2 + V(\phi)\right). \tag{6}$$

Here dot and prime denote differentiation with respect to cosmic time and $\phi$ respectively. We assume that inflation has already provided us with a flat universe by the time the largest observable scales cross the Hubble radius.

By differentiating Eq. (6) with respect to $t$ and substituting in Eq. (5), we arrive at the "momentum" equation

$$2\dot{H} = -\kappa^2 \dot{\phi}^2. \tag{7}$$

All minimal slow-roll models are examples of sub-inflationary behavior, which is defined by the condition $\dot{H} < 0$. Super-inflation, where $\dot{H} > 0$, cannot occur here, though it is possible in more complex scenarios [6, 7]. We may divide both sides of this equation by $\dot{\phi}$ if this quantity does not pass through zero. This allows us to eliminate the time-dependence in the Friedmann equation [Eq. (6)] and derive the first-order, non-linear (Hamilton-Jacobi) differential equations

$$(H')^2 - \frac{3}{2}\kappa^2 H^2 = -\frac{1}{2}\kappa^4 V(\phi); \qquad \kappa^2 \dot{\phi} = -2H'. \tag{8}$$

We now consider the production of density perturbations that arise as the result of quantum-mechanical fluctuations of fields in de Sitter space. First, let's consider scalar density fluctuations produced if we assume that the inflaton field $\phi$ is a massless, minimally coupled field. (Later we will include the corrections due to the fact that the inflaton field has a potential.)

Just as fluctuations in the density field may be expanded in a Fourier series as in Eq. (1), the fluctuations in the inflaton field may be expanded in terms of its Fourier coefficients $\delta\phi_\mathbf{k}$: $\delta\phi(\mathbf{x}) \propto \int \delta\phi_\mathbf{k} \exp(-i\mathbf{k}\cdot\mathbf{x})d^3k$. During inflation there is an event horizon as in de Sitter space, and quantum-mechanical fluctuations in the Fourier components of the inflaton field are given by [8]

$$k^3 |\delta\phi_\mathbf{k}|^2/2\pi^2 = (H/2\pi)^2, \tag{9}$$

where $H/2\pi$ plays a role similar to the Hawking temperature of black holes. Thus, when a given mode of the inflaton field leaves the Hubble radius during inflation, it has impressed upon it quantum mechanical fluctuations. In analogy to Eq. (3), what is called the fluctuation in the inflaton field on scale $k$ is proportional to $k^{3/2}|\delta\phi_\mathbf{k}|$, which by Eq. (9) is proportional to $H/2\pi$. Fluctuations in $\phi$ lead to perturbations in the energy density:

$$\delta\rho_\phi = \delta\phi(\partial V/\partial\phi). \tag{10}$$

Now considering the fluctuations as a particular mode leaves the Hubble radius during inflation, we may construct the gauge invariant quantity $\zeta$ using the fact that during inflation $\rho_0 + p_0 = \dot{\phi}^2$:

$$\zeta = \delta\phi\left(\frac{\partial V}{\partial \phi}\right)\frac{1}{\dot{\phi}^2}. \tag{11}$$

Now using Eqs. (8), the amplitude of the density perturbation when it crosses the Hubble radius *after* inflation is

$$\left(\frac{\delta\rho}{\rho}\right)^{\text{HOR}}_\lambda \equiv \frac{m}{\sqrt{2}}A_S(\phi) = \frac{m\kappa^2}{8\pi^{3/2}}\frac{H^2(\phi)}{|H'(\phi)|} \propto \frac{V^{3/2}(\phi)}{m_{Pl}^3 V'(\phi)}, \tag{12}$$

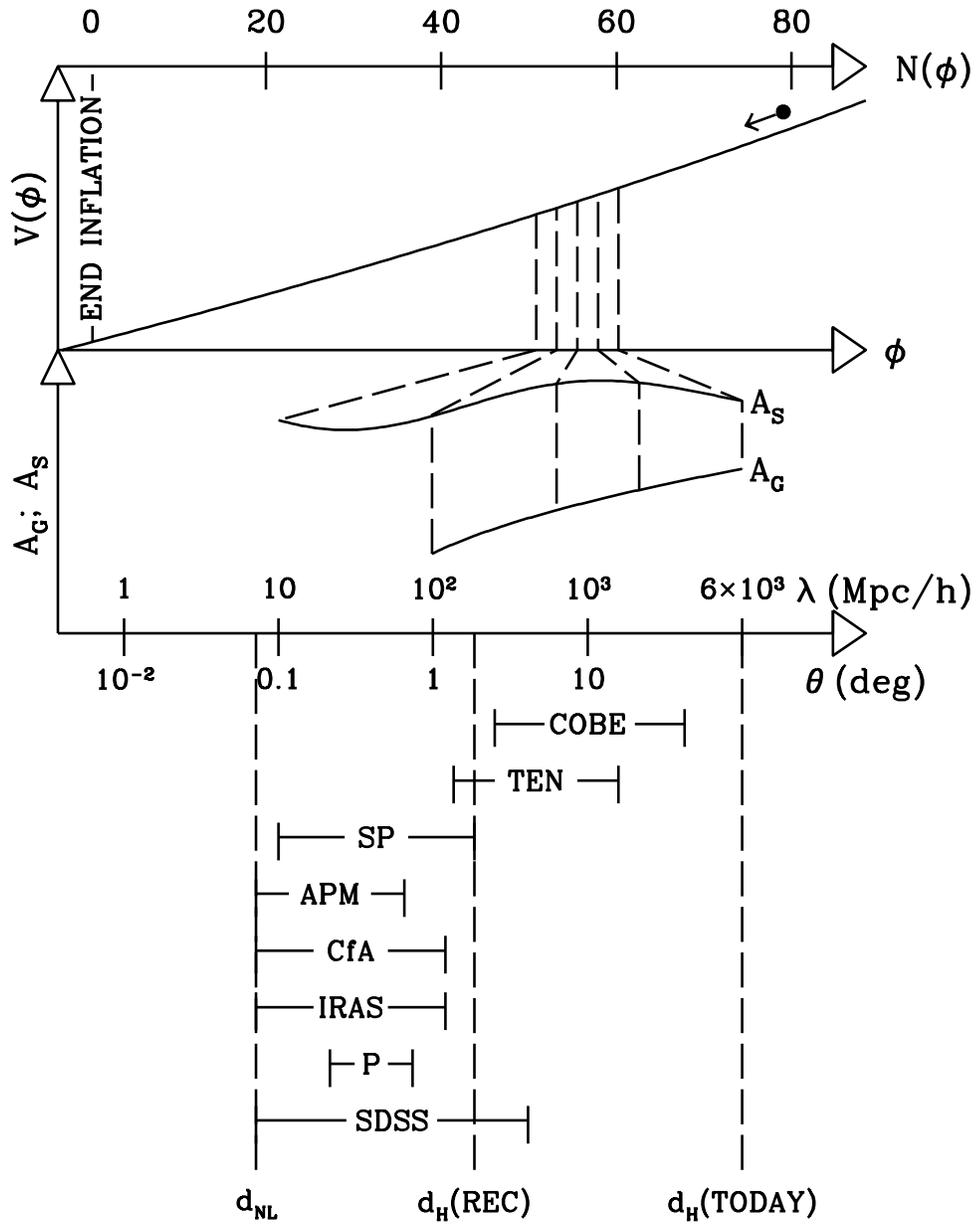

Fig. 1: The basic idea of inflation is that as the field evolves in the potential, quantum fluctuations in the inflaton field produce scalar density perturbations $A_S$, while fluctuations in the transverse, traceless metric components produce tensor gravitational wave perturbations, $A_G$. For reconstruction the two main steps involve converting the observations (lower half of figure) into the primordial scalar ($A_S$) and tensor ($A_G$) fluctuation spectra and then working in reverse to reconstruct the potential $V(\phi)$. The main observational information from the cosmic microwave background arises through the Cosmic Background Explorer (COBE) satellite [9], and the Tenerife (TEN) [10] and South Pole (SP) [11] collaborations. Galaxy surveys (APM [12], CfA [13], IRAS [14,15]) may provide useful information up to $100h^{-1}$ Mpc, while the Sloan Digital Sky Survey (SDSS) [16] should extend to the lowest scales measured by COBE. Peculiar velocity measurements using the POTENT (P) [17] methods are important on intermediate scales. The angle $\theta$ measures angular scales on the CMBR in degrees, and length scales $\lambda$ are in units of $h^{-1}$ Mpc. $d_H$ refers to the horizon size today and at recombination and $d_{\rm NL} \approx 8h^{-1}$ Mpc is the scale of non-linearity. Perfect observations will only reconstruct a small portion of the inflaton potential corresponding to between $53 \leq \Delta N \leq 60$ e-foldings before the end of inflation.

where $H(\phi)$ and $H'(\phi)$ are to be evaluated when the scale $\lambda$ crossed the Hubble radius *during* inflation. The constant $m$ equals 2/5 or 4 if the perturbation re-enters during the matter or radiation dominated eras respectively.[2] Now we wish to know the $\lambda$-dependence of $(\delta\rho/\rho)_\lambda$, while the right-hand side of the equation is a function of $\phi$ when $\lambda$ crossed the Hubble radius during inflation. We may find the value of the scalar field when the scale $\lambda$ goes outside the Hubble radius in terms of the number of *e*-foldings of growth in the scale factor between Hubble radius crossing and the end of inflation.

It is quite a simple matter to calculate the number of *e*-foldings of growth in the scale factor that occur as the scalar field rolls from a particular $\phi$ to the end of inflation $\phi_e$:

$$N(\phi) \equiv \int_{t_e}^{t} H(t')dt' = -\frac{\kappa^2}{2}\int_{\phi}^{\phi_e}\frac{H(\phi')}{H'(\phi')}d\phi'. \tag{13}$$

The total amount of inflation is given by $N_{\rm TOT} \equiv N(\phi_i)$, where $\phi_i$ is the initial value of $\phi$ at the start of inflation (when $\ddot{a}$ first becomes positive). In general, the number of *e*-folds between when a length scale $\lambda$ crossed the Hubble radius during inflation and the end of inflation is given by [18]

$$N(\lambda) = 45 + \ln(\lambda/{\rm Mpc}) + \frac{2}{3}\ln(M/10^{14}\,{\rm GeV}) + \frac{1}{3}\ln(T_{\rm RH}/10^{10}\,{\rm GeV}), \tag{14}$$

where $M$ is the mass scale associated with the potential and $T_{\rm RH}$ is the "re-heat" temperature. Relating $N(\lambda)$ and $N(\phi)$ from Eq. (13) results in an expression between $\phi$ and $\lambda$.

In addition to the scalar density perturbations caused by de Sitter fluctuations in the inflaton field, there are gravitational mode perturbations, $g_{\mu\nu} \to g_{\mu\nu}^{\rm FRW} + h_{\mu\nu}$, caused by de Sitter fluctuations in the metric tensor [19,20]. Here, $g_{\mu\nu}^{\rm FRW}$ is the Friedmann–Robertson–Walker metric and $h_{\mu\nu}$ are the metric perturbations. That de Sitter space fluctuations should lead to fluctuations in the metric tensor is not surprising, since after all, gravitons are the propagating modes associated with transverse, traceless metric perturbations, and they too behave as minimally coupled scalar fields. The dimensionless tensor metric perturbations can be expressed in terms of two graviton modes we will denote as $h$. Performing a Fourier decomposition of $h$, $h(\vec{\mathbf{x}}) \propto \int \delta h_k \exp(-i\vec{\mathbf{k}}\cdot\vec{\mathbf{x}})d^3k$, we can use the formalism for scalar field perturbations simply by the identification $\delta\phi_{\mathbf{k}} \to h_{\mathbf{k}}/\kappa\sqrt{2}$, with resulting quantum fluctuations [cf. Eq. (9)]

$$k^3|h_{\mathbf{k}}|^2/2\pi^2 = 2\kappa^2(H/2\pi)^2. \tag{15}$$

While outside the Hubble radius, the amplitude of a given mode remains constant, so the amplitude of the dimensionless strain on scale $\lambda$ when it crosses the Hubble radius after inflation is given by

$$\left|k^{3/2}h_{\mathbf{k}}\right|_\lambda^{\rm HOR} \equiv A_G(\phi) = \frac{\kappa}{4\pi^{3/2}}\,H(\phi) \sim \frac{V^{1/2}(\phi)}{m_{Pl}^2}, \tag{16}$$

where once again $H(\phi)$ is to be evaluated when the scale $\lambda$ crossed the Hubble radius *during* inflation.

---

[2]The 4 for radiation is appropriate to the uniform Hubble constant gauge. One occasionally sees a value 4/9 instead which is appropriate to the synchronous gauge. The matter domination factor is the same in either case. Note also that it is exact for matter domination, but for radiation domination it is only strictly true for modes much larger than the Hubble radius, and there will be corrections in the extrapolation down to the size of the Hubble radius.

# RECONSTRUCTION EQUATIONS TO SECOND ORDER

To some extent all inflationary calculations rely on the use of the slow-roll approximation. In the form we present here, the slow-roll approximation is an expansion in terms of quantities defined from derivatives of the Hubble parameter $H$. In general there are an infinite hierarchy of these which can in principle all enter at the same order in an expansion.

The slow-roll approximation arises in two separate places. The first is in simplifying the classical inflationary dynamics of expansion, with the lowest-order approximation ignoring the contribution of the inflaton's kinetic energy to the expansion rate. The second is in the calculation of the perturbation spectra; the standard expressions are true only to lowest-order in slow-roll. In the expressions in the previous section, we utilized the Hamilton-Jacobi approach [21] to treat the dynamical evolution exactly.

A very elegant calculation of the perturbation spectra to next order in slow-roll has now been provided by Stewart and Lyth [22]. The slow-roll approximation can be specified by parameters defined from derivatives of $H(\phi)$. There are in general an infinite number of these as each derivative is independent, but usually only the first few enter into any expressions. We shall require the first two, which are all of the same order when defined by

$$\epsilon(\phi) = \frac{2}{\kappa^2} \left[\frac{H'(\phi)}{H(\phi)}\right]^2 ; \qquad \eta(\phi) = \frac{2}{\kappa^2} \frac{H''(\phi)}{H(\phi)} . \tag{17}$$

The slow-roll approximation applies when these slow-roll parameters are small in comparison to unity. The condition for inflation, $\ddot{a} > 0$, is precisely equivalent to $\epsilon < 1$.

The lowest-order expressions for the scalar ($A_S$) and tensor ($A_G$) amplitudes assume $\{\epsilon, \eta\}$ are negligible compared to unity. Improved expressions for the scalar and tensor amplitudes for finite but small $\{\epsilon, \eta\}$ were found by Stewart and Lyth [22]:

$$\begin{aligned} A_S &\simeq -\frac{\sqrt{2}\kappa^2}{8\pi^{3/2}} \frac{H^2}{H'} \left[1 - (2C+1)\epsilon + C\eta\right] \\ A_G &\simeq \frac{\kappa}{4\pi^{3/2}} H \left[1 - (C+1)\epsilon\right] , \end{aligned} \tag{18}$$

where $C = -2 + \ln 2 + \gamma \simeq -0.73$ is a numerical constant, $\gamma \approx 0.577$ being the Euler constant. The right hand sides of these expressions are evaluated when the scale in question crosses the Hubble radius during inflation, $2\pi/\lambda = aH$. The spectra can equally well be considered to be functions of wavelength or of the scalar field value.

The standard results to lowest-order are given by setting the square brackets to unity. Historically it has been common even for this result to be written as only an approximate equality (the ambiguity arising primarily because of a vagueness in defining the precise meaning of the density perturbation), though the precise normalization to lowest-order was established some time ago by Lyth [23] (see also the discussion in [2]).

The improved expressions for the spectra in Eqs. (18) are accurate in so far as $\epsilon$ and $\eta$ are sufficiently slowly varying functions that they can be treated adiabatically as constants while a given scale crosses outside the Hubble radius. Corrections to this would enter at next order. This differs from the usual situation in which $H$ is treated adiabatically. For the standard calculation to be strictly valid $H$ must be constant, but provided it varies sufficiently slowly (characterized by small $\epsilon$ and $|\eta|$), it can be evaluated separately at each epoch. This injects a scale dependence into the spectra. There is a special case corresponding to power-law inflation for which $\epsilon$ and $\eta$ are precisely constant and equal to each other. In this case there are exact expressions for the perturbation [22,24]. Furthermore, the corrections to each spectrum are the same and they cancel when the ratio is taken. In the general case $\epsilon$ and $\eta$ may be treated as different constants if it is assumed that the timescale for their evolution is much longer than the timescale for perturbations to be imprinted on a given scale. This assumption worsens as $\eta$ is removed from $\epsilon$, which would be characterized by the next order terms becoming large.

It is useful to define the dimensionless quantities $\widetilde{\phi}$ and $v$, and a dimensionless derivative denoted by a dot:

$$\widetilde{\phi} \equiv \frac{\kappa}{\sqrt{2}}\phi; \qquad v(\widetilde{\phi}) \equiv \frac{\kappa^4}{48\pi^3}V(\phi) \, ; \qquad \dot{X} \equiv \frac{dX}{d\widetilde{\phi}} \, . \tag{19}$$

In addition, we can use the identity $\eta = \epsilon + \dot{\epsilon}/2\sqrt{\epsilon}$ and adopt as the expansion variables $\sum_{n=0} \epsilon^{-n/2} d^n \epsilon / d\widetilde{\phi}^n$. In terms of these variables, the expressions for $A_S(\lambda)$, $A_G(\lambda)$, $v$, and the $\widetilde{\phi}$–$\lambda$ relation become

$$\begin{aligned}
A_G &= \frac{\kappa}{4\pi^{3/2}} H \left[1 - (C+1)\epsilon\right] \\
A_S &= -\frac{\kappa}{4\pi^{3/2}} H \frac{1}{\sqrt{\epsilon}} \left[1 - (C+1)\epsilon + \frac{C}{2}\frac{\dot{\epsilon}}{\sqrt{\epsilon}}\right] \\
v &= A_G^2 \left[1 + \left(2C + \frac{5}{3}\right)\epsilon\right] = A_S^2 \epsilon \left[1 + \left(2C + \frac{5}{3}\right)\epsilon - C\frac{\dot{\epsilon}}{\sqrt{\epsilon}}\right] \\
\frac{d\widetilde{\phi}}{d\lambda} &= \frac{\sqrt{\epsilon}}{\lambda}(1 + \epsilon).
\end{aligned} \tag{20}$$

In the third expression, $v$ depends upon $A_G(\lambda)$ and $\epsilon$. Since we anticipate that we will only have information about $A_G(\lambda)$ at the largest scales, we have to use the "consistency" equation (also called the evolution equation) to relate $A_G(\lambda)$ to the more experimentally accessible $A_S(\lambda)$ at the expense of introducing the additional $\dot{\epsilon}/\sqrt{\epsilon}$ term.[3] This was done through the identity

$$\frac{A_G^2}{A_S^2} = \epsilon \left[1 - C\frac{\dot{\epsilon}}{\sqrt{\epsilon}}\right] . \tag{21}$$

which follows from the expressions for $A_S$ and $A_G$ in Eq. (20). Now we develop the evolution equation by taking the derivative of $A_S(\phi)$:

$$\begin{aligned}
\frac{\dot{A_S}}{A_S} &= \frac{1}{\epsilon^{1/2}} \left(\epsilon - \frac{1}{2}\frac{\dot{\epsilon}}{\sqrt{\epsilon}}\right) + \frac{1}{\epsilon^{1/2}} \left[\frac{C}{2}\frac{\ddot{\epsilon}}{\epsilon} - (C+1)\frac{\dot{\epsilon}}{\sqrt{\epsilon}}\epsilon - \frac{C}{4}\left(\frac{\dot{\epsilon}}{\sqrt{\epsilon}}\right)^2\right] \\
&\quad \times \left[1 + (C+1)\epsilon - \frac{C}{2}\frac{\dot{\epsilon}}{\sqrt{\epsilon}}\right] .
\end{aligned} \tag{22}$$

Now in addition to the expansion in $\epsilon$ and its derivatives, a truncation is necessary. The truncation here is to assume $\ddot{\epsilon}/\epsilon \ll \dot{\epsilon}/\sqrt{\epsilon}$. With this truncation, to second order

$$\frac{\dot{A_S}}{A_S} = \frac{1}{\epsilon^{1/2}} \left[\epsilon - \frac{1}{2}\frac{\dot{\epsilon}}{\sqrt{\epsilon}}\right] . \tag{23}$$

Now we can express $\dot{A_S}/A_S$ in terms of the spectral index $1 - n \equiv d\ln A_S^2 / d\ln \lambda$, and the evolution equation becomes

$$\begin{aligned}
\frac{\dot{\epsilon}}{\sqrt{\epsilon}} &= 2\epsilon - (1-n) \\
\frac{d\epsilon}{d\lambda} &= \frac{\epsilon}{\lambda}(1+\epsilon)\left[2\epsilon - (1-n)\right],
\end{aligned} \tag{24}$$

---

[3]Of course if the consistency equation is only used to evolve $\epsilon$, it can not be used as a check of inflation as discussed in [3].

where for the second equality we have used the $d\widetilde{\phi}/d\lambda$ expression. This evolution equation serves two purposes. It removes $\dot{\epsilon}/\sqrt{\epsilon}$ from the equation for $v$, and it is a differential equation that can be evolved to give $\epsilon$ as a function of $\lambda$. To solve the equation it is necessary to know the spectral index as a function of $\lambda$, along with the initial condition $\epsilon(\lambda_0)$ as a function of $1 - n_0$, $A_G(\lambda_0)$ and $A_S(\lambda_0)$.

So the system to be solved can be expressed as

$$\begin{aligned} v[\widetilde{\phi}(\lambda)] &= A_S^2(\lambda)\epsilon(\lambda)\left\{1 + \frac{5}{3}\epsilon(\lambda) + C\left[1 - n(\lambda)\right]\right\} \\ \frac{d\widetilde{\phi}}{d\lambda} &= \frac{\epsilon^{1/2}(\lambda)}{\lambda}\left[1 + \epsilon(\lambda)\right] \\ \frac{d\epsilon}{d\lambda} &= \frac{\epsilon}{\lambda}(1 + \epsilon)\left[2\epsilon - (1 - n)\right]. \end{aligned} \quad (25)$$

In the next section we discuss the simplification of the above expressions obtained by dropping the second-order terms and working to first order. In the section after that we solve an example first-order problem.

## FIRST-ORDER APPROXIMATION TO RECONSTRUCTION

To first order in the slow-roll expansion variables the expressions simplify considerably. For example, to first order, $\epsilon = A_G^2/A_S^2$, $v = A_G^2$, and $\lambda d\widetilde{\phi}/d\lambda = A_G/A_S$. The evolution–consistency equation is also quite simple. It can be written as

$$\frac{\lambda}{A_G(\lambda)}\frac{dA_G(\lambda)}{d\lambda} = \frac{A_G^2(\lambda)}{A_S^2(\lambda)}. \quad (26)$$

Again, the procedure will be the same as in the second-order case. The potential depends upon $A_G(\lambda)$, about which we will have information only on the largest scales (possibly only on one scale), so we specify the initial value of $A_G(\lambda)$, and use the consistency–evolution equation to evolve $A_G(\lambda)$ in terms of $A_S(\lambda)$. We can thus express the system to be solved in terms of two equations and a single first-order differential equation which can easily be solved in terms of the initial value $A_G(\lambda_0)$, yielding:

$$\begin{aligned} v[\widetilde{\phi}(\lambda)] &= \left[A_G^{-2}(\lambda_0) - 2\int_{\lambda_0}^{\lambda}\frac{d\lambda'}{\lambda'}\frac{1}{A_S(\lambda')}\right]^{-1} \\ \pm\widetilde{\phi} &= \int_{\lambda_0}^{\lambda}\frac{d\lambda'}{\lambda'}\frac{v^{1/2}[\widetilde{\phi}(\lambda')]}{A_S(\lambda')} \end{aligned} \quad (27)$$

## A WORKED EXAMPLE

Let's assume a simple power-law potential of the form $V(\phi) = \lambda_\phi \phi^4$ with $\lambda_\phi = 4 \times 10^{-14}$. This generates perturbation spectra of the form (evaluated at horizon crossing after inflation)

$$\begin{aligned} A_S(\lambda) &= 4 \times 10^{-8}\left[50 + \ln(\lambda/\lambda_0)\right]^{3/2} \\ A_G(\lambda) &= 4 \times 10^{-8}\left[50 + \ln(\lambda/\lambda_0)\right]. \end{aligned} \quad (28)$$

On any scale, the number of statistically independent sample measurements of the spectra that can be made is finite. Given that the underlying inflationary fluctuations are stochastic, one obtains only a limited set of realizations from the complete probability distribution function. Such a subset may insufficiently specify the underlying distribution, which is the quantity predicted by an inflationary model. The cosmic variance is an important matter of principle, being a source of uncertainty which remains even if perfectly accurate experiments could be carried out. At any stage in the history of the Universe, it is impossible to specify accurately the properties (most significantly the variance, which is what the spectrum specifies assuming gaussian statistics) of the probability distribution function pertaining to perturbations on scales close to that of the observable Universe.

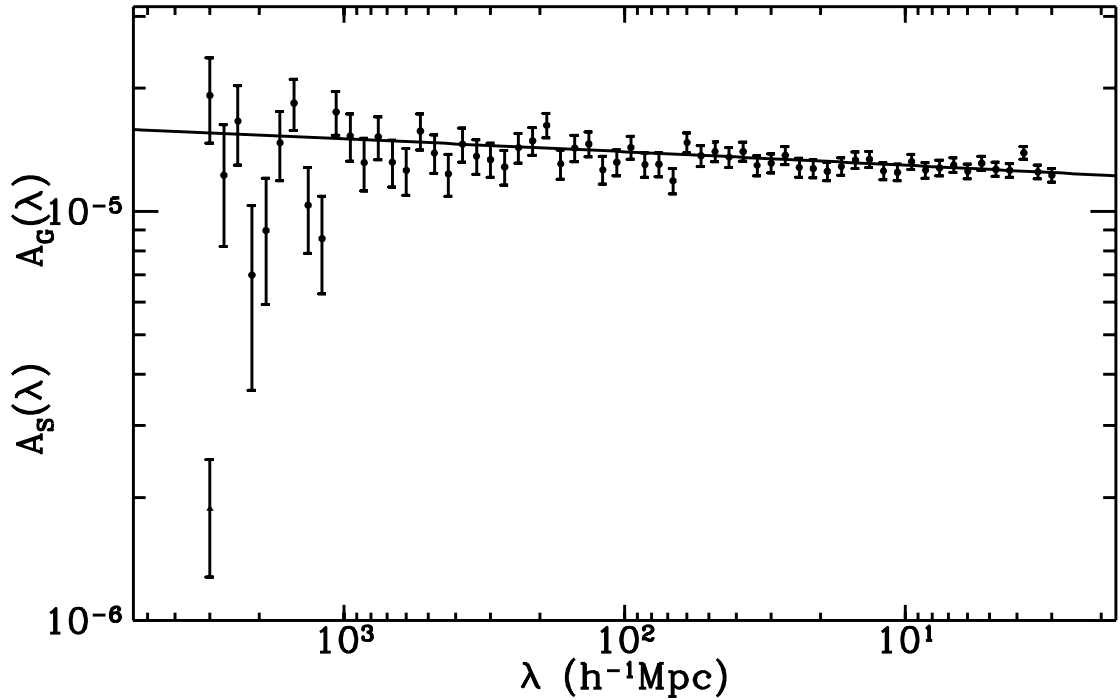

*Fig. 2: An illustration of an anticipated data set limited by cosmic variance. The data was generated with a $\lambda_\phi \phi^4$ potential with $\lambda_\phi = 4 \times 10^{-14}$. The upper points are $A_S(\lambda)$, while the single lower point is $A_G(\lambda)$. The solid line is the mean $A_S(\lambda)$, while the mean $A_G(\lambda_0)$ is $2 \times 10^{-6}$.*

Even assuming "perfect" observations, cosmic variance sets a lower limit on the uncertainty at any one scale. Assuming that the only errors come from cosmic variance, the determination of the spectra might look like in Fig. 2. In the realization generated by the random number generator, the value of $A_G(\lambda_0)$ is $1.87 \times 10^{-6}$, slightly below the ensemble mean of $2 \times 10^{-6}$.

As a first exercise, we simply perform a first-order reconstruction by doing a simple trapezoidal integration, and making the naïve assumption that the errors are uncorrelated. If we do that we obtain the reconstructed potential shown in Fig. 3. Also shown in Fig. 3 by the solid curve is the actual potential used to generate the synthetic data from which the potential was reconstructed.

There are several things we can notice in Fig. 3. First of all, reconstruction works: the true potential is within the error bars. The second obvious feature is that the slope of the reconstructed data is better than one might expect given the errors.

This feature can be explored by taking another approach to the uncertainty introduced in $A_G(\lambda_0)$ by cosmic variance. Let's ignore that error, and pick three realizations of $A_G(\lambda_0)$, one at the "measured" value, one $1\sigma$ above the measured value, and one $1\sigma$ below the measured value. (Here "$\sigma$" is the value determined by cosmic variance.) If we do that, we generate the three curves shown in Fig. 4. Although we can't tell which of the curves is the true potential, we know that the true potential is one of a family of curves bounded by the two extremes in the figure.

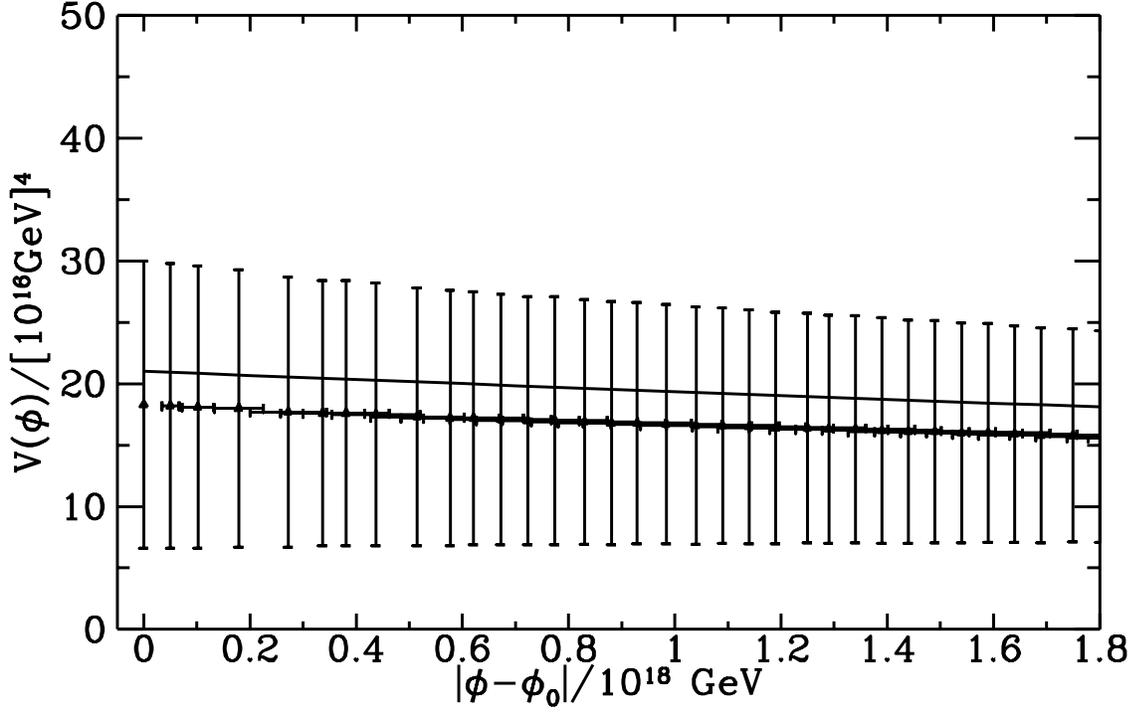

*Fig. 3: First-order reconstruction of the example $\lambda_\phi \phi^4$ potential. The solid line is the actual potential, while the points and associated errors were generated from the data of Fig. 2.*

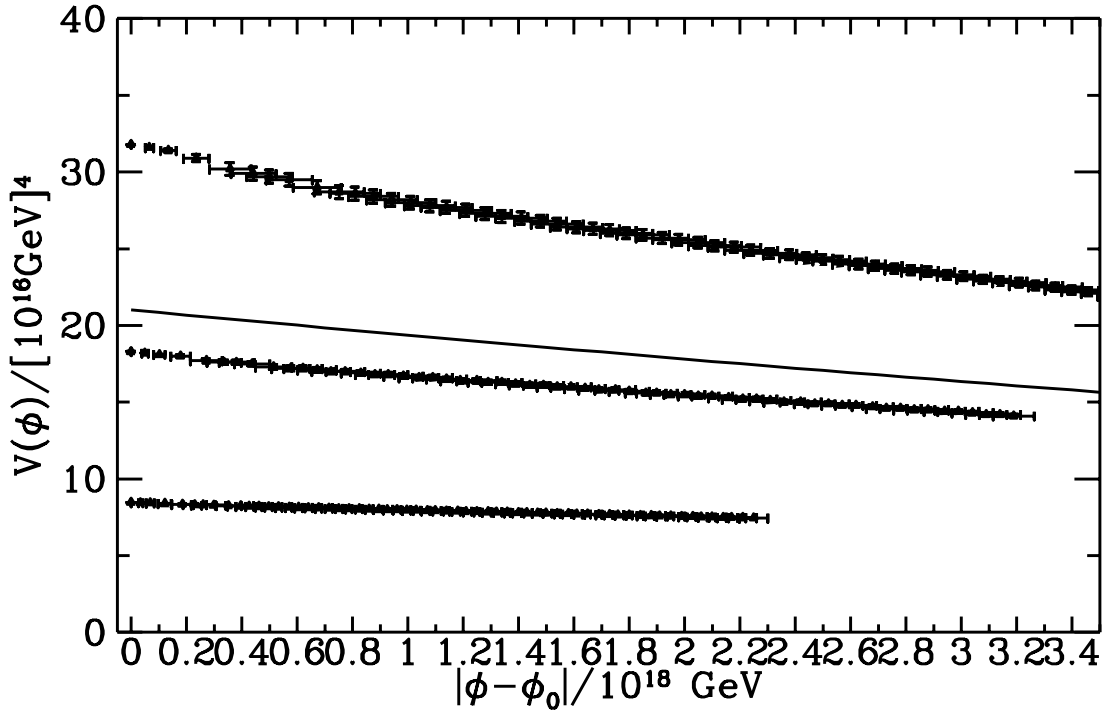

*Fig. 4: The reconstructed potential ignoring uncertainty in $A_G(\lambda_0)$ for three choices of $A_G(\lambda_0)$ corresponding to the midpoint and $\pm 1\sigma$.*

We can understand why this occurs, because if we look at the slope of $v^{-1}$, the initial value of $A_S(\lambda)$ drops out, and the contribution comes from adding together a large number of different $A_S(\lambda)$. Since we are combining a large number of data points, the central limit theorem tells us that the errors in the reconstructed potential will become small.

## CONCLUSIONS

The quantum-mechanical fluctuations impressed upon the metric during inflation depend upon the inflaton potential. During inflation the Hubble expansion takes microscopic fluctuations of wavelength of order $10^{-28}$cm and stretches them to super-Hubble-radius size where they are frozen. Today they appear on scales as large as the observable Universe, $10^{+28}$cm. It is possible to read the fossil record of the fluctuations by observing cosmic microwave background fluctuations and the power spectrum of large-scale structure.

If the tensor perturbations are large enough to be identified, and if the scalar power spectrum is determined, the inflaton potential may be reconstructed.

Hence cosmology and astrophysics may provide the first concrete piece of the potential of energy scales of $10^{16}$GeV or so.

## ACKNOWLEDGMENTS

EJC and JEL are supported by the PPARC. EWK and JEL are supported at Fermilab by the DOE and NASA under Grant NAGW–2381. ARL is supported by the Royal Society. We would like to thank David Lyth and Michael Turner for helpful discussions.

## REFERENCES

1. L. M. Krauss and M. White, Phys. Rev. Lett. **69**, 869 (1992); R. L. Davis, H. M. Hodges, G. F. Smoot, P. J. Steinhardt and M. S. Turner, Phys. Rev. Lett. **69**, 1856 (1992); A. R. Liddle and D. H. Lyth, Phys. Lett. **291B**, 391 (1992); J. E. Lidsey and P. Coles, Mon. Not. R. astr. Soc. **258**, 57P (1992); D. S. Salopek, Phys. Rev. Lett. **69**, 3602 (1992); F. Lucchin, S. Matarrese, and S. Mollerach, Astrophys. J. Lett. **401**, 49 (1992); T. Sourdeep and V. Sahni, Mod. Phys. Lett. A **7**, 3541 (1992).

2. A. R. Liddle and D. H. Lyth, Phys. Rep **231**, 1 (1993).

3. E. J. Copeland, E. W. Kolb, A. R. Liddle and J. E. Lidsey, Phys. Rev. Lett. **71**, 219 (1993); Phys. Rev. D**48**, 2529 (1993).

4. H. M. Hodges and G. R. Blumenthal, Phys. Rev. D**42**, 3329 (1990).

5. J. M. Bardeen, Phys. Rev. D**22**, 1882 (1980).

6. J. E. Lidsey, Phys. Lett. **273B**, 42 (1991).

7. F. Lucchin and S. Matarrese, Phys. Rev. D**32**, 1316 (1985); Phys. Lett. **164B**, 282 (1985).

8. T. Bunch and P. C. W. Davies, *Proc. Roy. Soc. London* **A360**, 117 (1978).

9. G. F. Smoot *et al.*, Astrophys. J. Lett. **396**, L1 (1992); E. L. Wright *et al.*, Astrophys. J. Lett. **396**, L13 (1992).

10. A. A. Watson *et al*, *Nature* **357**, 660 (1992).

11. T. Gaier, J. Schuster, J. Gunderson, T. Koch, M. Seiffert, P. Meinhold and P. Lubin, Astrophys. J. Lett. **398**, L1 (1992).

12. S. J. Maddox, G. Efstathiou, W. J. Sutherland and J. Loveday, *Mon. Not. R. astr. Soc.* **242**, 43p (1990).

13. M. J. Geller and J. P. Huchra, *Science* **246**, 879 (1989); M. Ramella, M. J. Geller and J. P. Huchra, Astrophys. J. **384**, 396 (1992).

14. W. Saunders *et al*, *Nature* **349**, 32 (1991); N. Kaiser, G. Efstathiou, R. Ellis, C. Frenk, A. Lawrence, M. Rowan-Robinson and W. Saunders, *Mon. Not. R. astr. Soc.* **252**, 1 (1991).

15. K. B. Fisher, M. Davis, M. A. Strauss, A. Yahil and J. P. Huchra, Astrophys. J. **389**, 188 (1992).

16. J. E. Gunn and G. R. Knapp, "The Sloan Digital Sky Survey," Princeton preprint POP-488 (1992); R. G. Kron in *ESO Conference on Progress in Telescope and Instrumentation Technologies*, ESO conference and workshop proceeding no. 42, p635, ed. M.-H. Ulrich (1992).

17. E. Bertschinger and A. Dekel, Astrophys. J. Lett. **336**, L5 (1989); A. Dekel, E. Bertschinger and S. M. Faber, Astrophys. J. **364**, 349 (1990); E. Bertschinger, A. Dekel, S. M. Faber, A. Dressler and D. Burstein, Astrophys. J. **364**, 370 (1990).

18. P. J. Steinhardt and M. S. Turner, Phys. Rev. D**29**, 2162 (1984); E. W. Kolb and M. S. Turner, *The Early Universe*, (Addison-Wesley, New York, 1990).

19. V. A. Rubakov, M. Sazhin, and A. Veryaskin, Phys. Lett. **115B**, 189 (1982); R. Fabbri and M. Pollock, Phys. Lett. **125B**, 445 (1983); B. Allen, Phys. Rev. D **37**, 2078 (1988); and L. Abbott and M. Wise, Nucl. Phys. **B244**, 541 (1984).

20. A. A. Starobinsky, *Sov. Astron. Lett.* **11**, 133 (1985).

21. D. S. Salopek and J. R. Bond, Phys. Rev. D**42**, 3936 (1990); J. E. Lidsey, Phys. Lett. **273B**, 42 (1991).

22. E. D. Stewart and D. H. Lyth, Phys. Lett. **302B**, 171 (1993).

23. D. H. Lyth, Phys. Rev. D**31**, 1792 (1985).

24. D. H. Lyth and E. D. Stewart, Phys. Lett. **274B**, 168 (1992).